\documentclass[aps,pra,twocolumn,groupedaddress,showpacs]{revtex4}

\usepackage{graphicx}

\begin{document}

\title{Interference of Bose-Einstein Condensates on an Atom Chip}

\author{Y. Shin}
\author{C. Sanner}
\author{G.-B. Jo}
\author{T.~A. Pasquini}
\author{M. Saba}
\author{W. Ketterle}
\author{D.~E. Pritchard}
\homepage[URL: ]{http://cua.mit.edu/ketterle_group/}
\affiliation{MIT-Harvard Center for Ultracold Atoms, Research
Laboratory of Electronics, Department of Physics, Massachusetts
Institute of Technology, Cambridge, Massachusetts, 02139}
\author{M. Vengalattore}
\author{M. Prentiss}
\affiliation{MIT-Harvard Center for Ultracold Atoms, Jefferson
Laboratory, Physics Department, Harvard University, Cambridge,
Massachusetts, 02138}

\date{\today}

\begin{abstract}
We have used a microfabricated atom chip to split a single
Bose-Einstein condensate of sodium atoms into two spatially
separated condensates. Dynamical splitting was achieved by
deforming the trap along the tightly confining direction into a
purely magnetic double-well potential. We observed the matter wave
interference pattern formed upon releasing the condensates from
the microtraps. The intrinsic features of the quartic potential at
the merge point, such as zero trap frequency and extremely high
field-sensitivity, caused random variations of the relative phase
between the two split condensates. Moreover, the perturbation from
the abrupt change of the trapping potential during the splitting
was observed to induce vortices.
\end{abstract}


\maketitle

Coherent manipulation of matter waves is the ultimate goal of
atom optics, and diverse atom optical elements have been
developed such as mirrors, beamsplitters, gratings, and
waveguides. An atom chip integrates these elements on a
microfabricated device allowing precise and stable
alignment~\cite{firstchip}. Recently, this atom chip technology
has been combined with Bose-Einstein condensed
atoms~\cite{OFS01,HHH01}, and opened the prospect for chip-based
atom interferometers with Bose-Einstein condensates. Despite
various technical problems~\cite{LCK02,FOK02,LSC03,JVS03,EAS04},
there have been advances toward that goal, such as excitationless
propagation in a waveguide~\cite{LCK02} and demonstration of a
Michelson interferometer involving splitting along the axis of a
single waveguide~\cite{WAB04}.

Coherent splitting of matter waves into spatially separate atomic
wave packets with a well-defined relative phase is a prerequisite
for further applications such as atom interferometry and quantum
information processing, and it has been a major experimental
challenge. The methods envisioned for coherent splitting on atom
chips can be divided in two classes. One is splitting in momentum
space and subsequently generating a spatial separation, using
scattering of atoms from a periodic optical
potential~\cite{MOM88,WAB04}. The other is dynamical splitting by
directly deforming a single wave packet into two spatially
separated wave packets, which can be considered as cutting off the
link between two wave packets, i.e., stopping tunneling through
the barrier separating two wave packets. Splitting in momentum
space has led to remarkably clean interferometric measurements
when the atoms were allowed to propagate freely after splitting,
but it has been pointed out that momentum splitting of confined
atoms (e.g. inside a waveguide) is problematic due to spatially
dependent phase shifts induced by atom-atom interactions during
separation~\cite{WAB04,OLD05}. Dynamical splitting in real space
instead is perfectly compatible with keeping atoms confined, a
feature beneficial to the versatility of interferometers. There
has been a theoretical debate concerning the adiabatic condition
for coherent dynamical splitting~\cite{JW97,LS98,MAC01,PSB04}. In
our recent experiment with an optical double-well potential, we
demonstrated that it is possible to dynamically split a
condensate into two parts in a coherent way~\cite{SSP04}.

In this work, we studied the dynamical splitting of condensates in
a purely magnetic double-well potential on an atom chip. We
developed an atom chip to generate a symmetric double-well
potential and succeeded in observing the matter wave interference
of two split condensates, from which the coherence of the
splitting process was investigated. We found that the mechanical
perturbations during splitting are violent enough to generate
vortices in condensates. We discuss the adiabatic condition of
the splitting process.

\begin{figure}
\begin{center}
\includegraphics{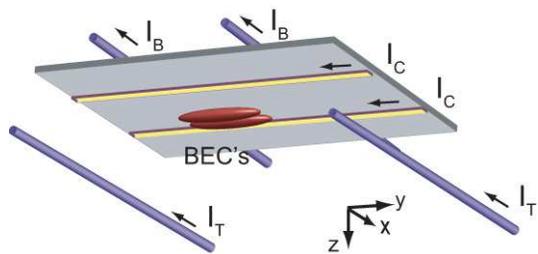}
\caption{Schematic diagram of the atom chip. A magnetic
double-well potential is created by two chip wires with a current
$I_C$ in conjunction with an external magnetic field. The
distance between the two chip wires was 300~$\mu$m. A pair of
external wires with $I_B$ provided the axial confinement along
the $y$ direction, and another pair of external wires with $I_T$
were used for reducing the anti-symmetry effect. (for details, see
text.) Gravity was in the $+z$ direction.\label{f:setup}}
\end{center}
\end{figure}

A magnetic double well potential was realized with an atom chip
using a two-wire scheme~\cite{HVB01}. The experimental setup of
the atom chip is shown in Fig.~\ref{f:setup}. When two chip wires
have currents, $I_C$, in the $-y$ direction and the external
magnetic field, $B_x$, is applied in the $+x$ direction, two lines
of local minima in the magnetic field are generated above the chip
surface. Each local minimum has a quadruple field configuration in
the $xz$ plane, and with an additional non-zero magnetic field in
the axial direction ($y$-direction), two Ioffe-Pritchard magnetic
traps can be formed. The relative magnitude of $B_x$ to the field
from $I_C$ determines the direction of separation and the
distance of the two traps. The atom chip was set to face downward
and the two traps are vertically (horizontally) separated when
$B_x<B_{x0}$ ($B_x>B_{x0}$). $B_{x0}=\mu_0 I_C/\pi d$ is the
critical field magnitude for merging two magnetic harmonic
potential to form a single quartic potential, where $d$ is the
distance between the two chip wires and $\mu_0$ is the
permeability of the vacuum. The merge point is located at the
middle of the two wires and $d/2$ away from the chip surface. In
our experiment, $d=300~\mu$m; thus, the splitting happened more
than 200~$\mu$m away from the chip wires to avoid deleterious
surface effects~\cite{LCK02,FOK02,LSC03,JVS03,EAS04}. The chip
wires of 12~$\mu$m height and 50~$\mu$m width were electroplated
with Au on a thermally oxidized Si substrate with a
2~$\mu$m-thick Au evaporated film. The chip was glued on an Al
block for heat dissipation~\cite{GKW04} and the current capacity
was 5~A in a continuous mode.

The axial trapping potential was carefully designed to ensure
that condensates split perpendicular to the axial direction and
stay in the same axial position. The two wells have opposite
responses to $B_z$: positive $B_z$ makes the left (right) well
move upward (downward). If $B_z$ changes along the axial
direction, the two wells are no longer parallel and the
gravitational force would cause an axial displacement of the two
split condensates. When endcap wires are placed only on the chip
surface as in our previous work~\cite{SSV04}, a non-zero field
gradient $\frac{\partial B_z}{\partial y}$ inevitably accompanies
a field curvature $\frac{\partial^2 B_y}{\partial y^2}$ for the
axial confinement, i.e., $B_z$ changes from positive to negative
along the axial direction. In order to provide the axial
confinement and at the same time minimize $\frac{\partial
B_z}{\partial y}$, we placed two pairs of external wires 1.5~mm
above and 4~mm below the chip surface. This three-dimensional
design of axial confinement was necessary for obtaining the
interference signal of two split condensates. Moreover,
maintaining the geometric symmetry of two wells will be crucial
for longer coherence time after splitting~\cite{SSP04}.

\begin{figure}
\begin{center}
\includegraphics{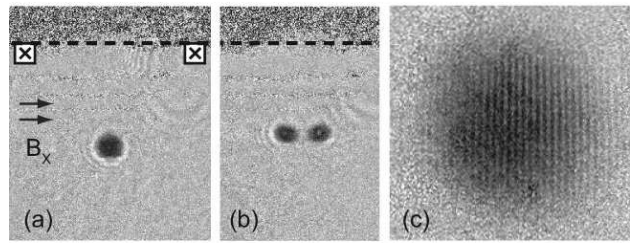}
\caption{Splitting of condensates. (a) Condensates are initially
loaded and prepared in the bottom well and (b) split into two
parts by increasing the external magnetic field, $B_x$. For
clarity, two condensates were split by 80~$\mu$m. The dash line
indicates the chip surface position. The currents in the chip
wires flow into the page and $B_x$ is parallel to the wire
separation. Two condensates were released from the magnetic
double-well potential and the matter wave interference pattern of
two condensates formed after time-of-flight. (c) Typical
absorption image of interference fringes taken after 22~ms
time-of-flight. The fringe spacing is 14.8~$\mu$m, corresponding
to a condensate separation of 25.8~$\mu$m. \label{f:typical}}
\end{center}
\end{figure}

The splitting process was demonstrated with the experimental
procedures described in Fig.~\ref{f:typical}. Bose-Einstein
condensates of $|F=1, m_F=-1\rangle$ $^{23}$Na atoms were
transferred and loaded in a magnetic trap generated by the atom
chip~\cite{LCK02,LGC02,SSV04}. Experimental parameters were
$I_C=1.8$~A, $B_{x0}=24$~G, $B_y=1$~G, and the axial trap
frequency $f_y=13$~Hz. Condensates were first loaded in the bottom
well, 500~$\mu$m away from the chip surface, brought up to
30~$\mu$m below the merge point in 1~s, and held there for 2~s to
damp out excitations. The long-living axial dipole excitation
induced in the transfer phase was damped by applying a repulsive
potential wall at the one end of the condensates with a
blue-detuned laser beam (532~nm)~\footnote{In a perfectly
symmetric double-well potential, two condensates would oscillate
in phase after splitting. Furthermore, this could be used for
developing a rotation-sensitive atom interferometer with a guiding
potential. However, the axial trap frequencies for the two wells
were found to be different by 12~$\%$ due to the imperfect
fabrication of wires.}. The whole procedures was carried out with
a radio-frequency (rf) shield and, just before splitting,
condensates contained over $8.0 \times 10^5$ atoms without a
discernible thermal population. Splitting was done by ramping
$\Delta B_x = B_x-B_{x0}$ linearly from -$140$~mG to $100\pm
20$~mG in 200~ms. The separation between two condensates was
controlled by the final value of $B_x$. The magnetic trap was
then quickly turned off within 20~$\mu$s, much shorter than the
inverse of any trap frequency, preventing random perturbations.
High-contrast matter wave interference fringes were observed after
releasing the condensates and letting them expand in
time-of-flight (Fig.~\ref{f:typical}), indicating that the
splitting procedure was smooth enough to produce two condensates
having uniform phases along their long axial axis perpendicular
to the splitting direction. In order to study the coherence of
the splitting, the relative phase of the two split condensates
was determined from the spatial phase of the matter wave
interference pattern.

\begin{figure}
\begin{center}
\includegraphics{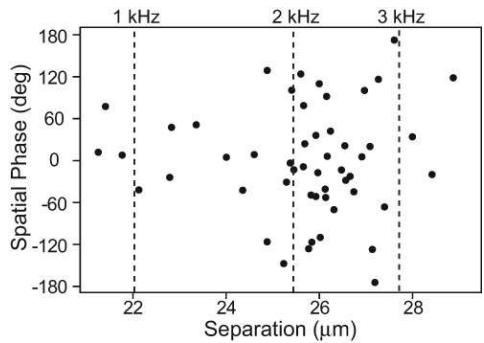}
\caption{Spatial phase of interference fringes. The separation of
two condensates was determined from the spacing of interference
fringes. Fifty repetitions of the same experiment are plotted. The
distance of the two trap centers is determined by the value of
the external magnetic field, $B_x$ when the atoms are released.
Three dash lines indicate the separations of two wells with the
barrier height of 1~kHz, 2~kHz, and 3~kHz,
respectively.\label{f:phase}}
\end{center}
\end{figure}

The relative phase of two split condensates turned out to be
unpredictable when they were fully separated
(Fig.~\ref{f:phase}). The separation of two condensates was
determined from the spacing, $\lambda_s$, of the interference
fringes, using the formula, $d=h t/m \lambda_s$ where $h$ is
Plank's constant, $m$ is atomic mass, and $t$ is time-of-flight.
The typical fringe spacing was $\lambda \approx 15$~$\mu$m with
$t=22$~ms, corresponding to $d\approx 26$~$\mu$m. Given the
precise knowledge of the fabricated wires, the full trap
parameters can be calculated. Assuming that the condensates
followed trap centers in the motional ground state, it was found
that when the barrier height was over 1.5~kHz, the relative phase
started to be random. Since the chemical potential of the
condensates, $\mu=1.4\pm 0.2$~kHz was very close to this barrier
height, the condensates just started to lose their coupling at
this point.

\begin{figure}
\begin{center}
\includegraphics{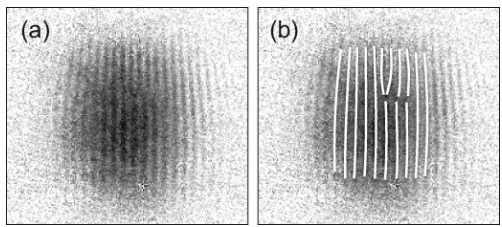}
\caption{Vortex interference. (a) An absorption image showing the
vortex interference pattern of a vortex state. The probability of
vortex generation was $\sim$8~$\%$ for the experimental
parameters of Fig.~\ref{f:phase}, where data points with vortices
were not included. Vortex interference patterns appeared more
frequently with faster splitting and further separation. (b) Same
as (a), but with lines indicating regions with constant
phase.\label{f:vortex}}
\end{center}
\end{figure}

Surprisingly, a phase singularity was observed in the interference
patterns with high visibility. The fork shape of interference
fringes represents a phase winding around a vortex
core~\cite{IGR01}. This vortex interference pattern appeared more
frequently with faster splitting and further separation. An
external perturbation can lead to internal excitations in
condensates. Splitting might be considered as slicing condensates
in two parts. The fact that the observed ``forks"
(Fig.~\ref{f:vortex}) always open towards the top implies that the
the slicing always occurred in the same direction and created
either vortices with positive charge on the left side or with
negative charge on the right side. A possible vortex formation
mechanism is topological imprinting when the zero point of the
magnetic field crosses though condensates resulting in a doubly
quantized vortex in spin-1 condensates~\cite{LGC02,SSV04}.
However, since we have never observed the interference pattern of
a doubly quantized vortex, we think that this scenario is
unlikely.

\begin{figure}
\begin{center}
\includegraphics{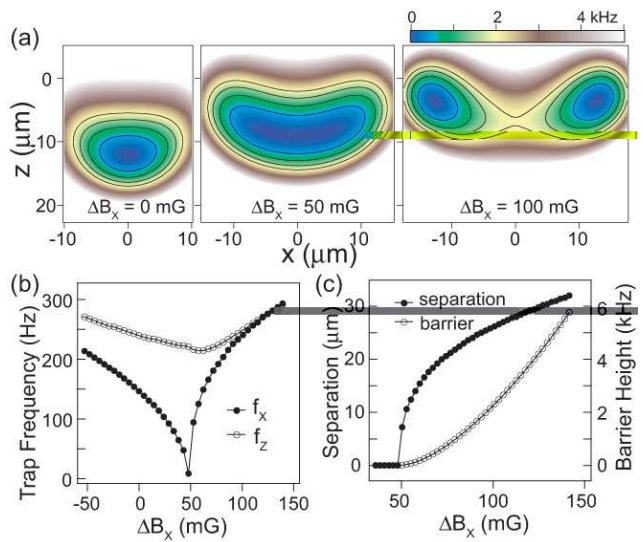}
\caption{Trapping potential during splitting. (a) Radial cross
sections of trapping potential including gravity for $\Delta
B_x$=0, 50, 100~mG, where $\Delta B_x$ is the field deviation
from the critical field magnitude $B_{x0}$ which is the field
magnitude for forming a single quartic trap. The origin of
coordinates is the merge point without gravity. Contour lines
correspond to 0.5, 1, 1.5, 2~kHz above the bottom of the trap. (b)
Trap frequencies in each direction. (c) Separation of two trap
centers and barrier height between two wells.\label{f:potential}}
\end{center}
\end{figure}

We now discuss how the trapping potential changes during the
splitting process (Fig.~\ref{f:potential}). When condensates split
into two wells, the trap frequency, $f_x$, in the splitting
direction vanishes and the separation of two wells abruptly
increases to 15~$\mu$m with a small magnetic field change of
$\delta B_x\approx 10$~mG. The quantity $\alpha \equiv
\frac{1}{f_x^2} \frac{\partial f_x}{\partial B_x} \frac{dB_x}{dt}$
parameterizes the external adiabaticity of the process for a
single particle, neglecting the collective excitations of a
condensate, and $\alpha \ll 1$ should be maintained to keep
condensates staying in the motional ground state. With
$\frac{dB_x}{dt}=1.2$~G/s, $\alpha < 1$ at $f_x>150$~Hz, but
obviously, $\alpha$ diverges to infinity near the merge point,
implying that the quartic potential with zero trap frequency makes
it impossible to satisfy the adiabatic condition during the whole
splitting process. The abrupt change of trapping potential will
induce mechanical perturbations of condensates. Subsequent
dissipation or coupling into internal excitation
modes~\cite{OFK03} would make the relative phase of two split
condensates unpredictable. The observed phase singularity
definitely shows the breakdown of adiabaticity.

One possible alternative to avoid passing through the merge point
is starting with two weakly-linked condensates in a double-well
potential where the barrier height is lower than the chemical
potential of condensates and controlling the coupling between two
condensates with a small change of the barrier height. This method
was used to reduce the motional perturbation in our previous
work~\cite{SSP04}. However, since the sensitivity of the trapping
potential to the magnetic field is extremely high when the trap
centers are close to the merge point, it was technically difficult
to have a stable double-well potential with a small barrier
height. The lifetime of condensates measured around the merge
point was $>5$~s away from the merge point ($\Delta B_x<-50$~mG or
$\Delta B_x>150$~mG) and $<100$~ms near the merge point
($0<\Delta B_x<100$~mG)~\footnote{For positions with $\Delta B_x
>0$ (``after" splitting), the condensates were moved to the left
well without passing through the merge point.}. With a barrier
height of 0.5~kHz in our experiment, the sensitivity of the
barrier height and the condensate separation to $B_x$ is
0.04~kHz/mG and 0.3~$\mu$m/mG, respectively. $\delta B_x=1$~mG
corresponds to $\delta I_C=7.5\times 10^{-5}$~A. Extreme current
stabilization and shielding of ambient magnetic field
fluctuations may be necessary for controlling a phase-coherent
splitting process. Another alternative for preparing a coherent
state of two spatially separated condensates is first preparing
two condensates in the ground states in each well and then
establishing a well-defined relative phase with an optical
method~\cite{SPS05}. This scheme is currently under investigation.

In conclusion, we demonstrated the interference of two
Bose-Einstein condensates released from an atom chip. The
condensates were created by dynamical splitting of a single
condensate and could be kept confined in a magnetic double well
potential, separated by an arbitrary distance. We studied the
coherence of the dynamical splitting process by measuring the
relative phase of two split condensates and identified technical
limitations, intrinsic to the magnetic field geometry, that
prevented coherent splitting with a predictable phase. This study
is a promising step in the route towards atom chip interferometers
and might serve as a guide for the design of future
microfabricated atom optics device.

This work was funded by ARO, NSF, ONR, DARPA and NASA. C.S.
acknowledges additional support from the Studienstiftung des
deutschen Volkes, G.-B. J. from the Samsung Lee Kun Hee
Scholarship Foundation, and M.S. from the Swiss National Science
Foundation.

\end{document}